# STRESS ORIENTATION EVALUATED FROM STRAIN LOCALISATION ANALYSIS IN AIGION FAULT

**Jean Sulem**[1]
CERMES, Ecole Nationale des Ponts et Chaussées, UR Navier
Marne-La-Vallée, France

## ABSTRACT

Within the frame of the 'CRL' (Corinth Rift Laboratory project) (Cornet et al, 2004a) centered on the south western sector of the Gulf of Corinth (http://www.corinth-rift-lab.org), fault zone cores from the active Aigion fault have been collected continuously from depths between 708 m to 782 m.
Inside this clayey core, a clear shearing surface with marked slip lines is visible on a plane that makes a 68° angle with respect to the core axis. This failure surface was not induced by the decompression process but is indeed a slip plane as clear striation is observed at the interface. On the basis of an elastoplastic constitutive model calibrated on triaxial tests on the clayey gouge, it is shown that shear band formation inside the clayey core is possible. The solution for the orientation of the shear band is compared to the orientation of an existing slip surface inside the clayey gouge and this result is used to deduce the orientation of the principal stresses. It is shown that as commonly observed in weak fault zones, the orientation of the principal stresses is locally almost parallel and perpendicular to the fault axis.

[1] Address : Jean Sulem
CERMES, Ecole Nationale des Ponts et Chaussées
6-8 Av. Blaise Pascal, Cité Descartes, Champs sur Marne
77455 Marne-La-Vallée Cedex 2 France
Phone :+33-(0)1 64 15 35 45 / Fax:+33 (0)1 64 15 35 62,
Email : sulem@cermes.enpc.fr

## INTRODUCTION

It is commonly observed in laboratory experiments as well as in field observations that the orientation of a failure plane (or fault surface) is controlled by the directions of the principal stresses. In the well established Mohr-Coulomb's theory, the inclination θ of the failure surface with respect to the direction of the minimum (in absolute value) principal stress is given as $\theta = \frac{\pi}{4} + \frac{\phi}{2}$ where ϕ is the friction angle of the material. For typical values of Coulomb friction angle ϕ, 30° to 50°, values of θ range from 60° to 70°, which is comparable with the range of observed failure plane inclinations. The milestone paper of Mohr (1900) was based on the observation of the regularity of slip lines or failure patterns in metals and the conviction that this regularity, as it is characterised by almost constant relative inclinations angles, can only be the result of material properties which prior to failure are the same everywhere in the tested specimens. Mohr-Coulomb's theory is commonly used in geomechanics as the dominant feature in the behaviour of geomaterials behaviour is its frictional character. The orientation of a failure surface can be deduced from the knowledge of the orientation of the principal stresses (not their magnitude) and of one material property (the internal friction angle). Inversely, the orientation of the principal stresses can be simply deduced, using Mohr-Coulomb's theory, from the orientation of the failure plane and the knowledge of the friction coefficient of the material.

Although the simplicity of this approach has made it very useful, the predictions of Mohr-Coulomb's theory have been criticized because it is commonly observed that the complete inelastic response of a material influences the conditions of incipient failure and not only one material parameter such as the internal friction angle. Failure of geomaterials is characterized by the formation and propagation of zones of localised shear deformation and the study and the modelling of strain localisation phenomena has proven to be very useful in the understanding of failure mechanisms. Based on the theoretical studies of material stability as developed by Hadamard (1903) for elastic materials and later extended by Thomas (1961), Hill (1962) and Mandel (1966) for inelastic materials, the localisation process is seen as an instability that can be predicted from the pre-failure constitutive behaviour of the material. The conditions for the onset of localisation are thus established by seeking the possible critical conditions for which the constitutive equations of the material (in the pre-localised stage) may allow the existence of a bifurcation point for which the deformation mode will localize into a planar band (Rice 1976, Vardoulakis, 1976). In this approach, the initiation of failure in the form of the incipient of a shear band is modelled as a constitutive instability and consequently a great number of studies have been dealing with the development of appropriate constitutive relationships which can predict satisfactorily the onset of shear banding. Geomaterials are characterised by a non-associated and non-coaxial plastic behaviour and this has important effects on the localisation process (Rudnicki and Rice, 1975).

In this paper, we propose to evaluate the orientation of the principal stresses acting in a fault zone, from a strain localisation analysis and from the comparison of the predicted shear band orientation to the orientation of a slip plane clearly observed inside the fault core. The analysis is based on an elastoplastic constitutive model calibrated on triaxial tests performed on samples extracted from the fault core. These samples have been collected at a depth of about 760m during the drilling of Aigion fault in the southern shore of the Gulf of Corinth, Greece, within the frame of the Corinth Rift Laboratory project.

# 1 GEOLOGICAL SETTING

The Gulf of Corinth is the most seismically active zone in Europe and the fastest opening area of continental break-up, with up to 1.5 cm/year of north–south extension and more than 1 mm/year of uplift of the southern shore of the Gulf (Tselentis and Makropoulos, 1986) (Fig. 1 from Cornet et al 2004a). Over the last years, the Gulf of Corinth has been a subject of intense research in the framework of the European multi-disciplinary cluster of projects called Corinth Rift Laboratory (CRL). The results of this research have been published in special issues of Comptes Rendus, Geosciences (2004) and Tectonophysics (2006). The general orientation of the rift is N 100°, and it is 105 km long. South of Aigion, the Gulf is limited by a series of sub-parallel normal faults dipping north with an angle in the range 55–70° (Fig. 2 from Moretti and al, 2003). Micarelli et al. (2003) published a detailed structural description of three faults of the studied area, carried out on natural outcrops of Helike and Pirgaki faults and on borehole cores from the Aigion fault. They report the existence of 2 m thick, highly-deformed cores. Around these cores, the host rocks present a clear damage zone. The total thickness of the damage zones varies from 10–20 m (Aigion) to 60 m (Pirgaki).

During drilling in the active Aigion fault, cuttings were collected, geophysical logs, including borehole imaging and sonic logs as well as vertical seismic profiles, were completed. This has provided material for a detailed lithological and structural investigation of the formations intersected by the well presented in the paper of Rettenmaier et al (2004). Results are summarized in Fig. 3 (from Cornet et al 2004b). Fault zone cores from the active Aigion fault have been collected continuously from depths between 708 m to 782 m.

At depth 760m the Aigion fault was intercepted, dipping at an angle of about 60°. The heart of the fault is a 1m-thick band made of clay-rich material derived from radiolarites (Fig. 4).

In recent papers results from mechanical laboratory analyses on specimens taken from the Aigion fault core have been presented (Sulem et al, 2004, 2005). Special attention has been paid on temperature effects on the behavior of the clayey core of the lower part of Aigion fault drilling core and applied to the analysis of shear heating mechanism during seismic slip (Sulem et al. 2006).

Inside the clayey core, a clear shearing surface with marked slip lines is visible on a plane that makes a 68° angle with respect to the core axis. This failure surface was not induced by the coring and extraction process but is indeed a slip plane as clear striation is observed at the interface (see Fig. 5). A strain localization analysis is performed to extract from the observation of this shear plane, information on the of state stress in the fault zone.

# 2 CONSTITUTIVE MODEL FOR AIGION CLAYEY GOUGE

## *2.1 Experimental results*

An elastoplastic constitutive model for Aigion clayey gouge has been proposed in previous work (Sulem et al. 2004) based on drained triaxial compression tests. For the sake of clarity, the main points are recalled here.

Drained triaxial compression tests have been performed at room temperature (22°C) at 8, 16 and 18 MPa confining pressure, on saturated samples of the clayey gouge. In the drained tests, a constant pore pressure $P_p$ of 0.5 MPa is applied to the samples.

For triaxial compression tests the axial and radial stresses are respectively $\sigma_z$ and $\sigma_r$, with $\sigma_r=\sigma_{confinement}$=constant.

The mobilized friction coefficient f and friction angle $\phi$ are expressed as (Vardoulakis and Sulem, 1995)

$$f = \frac{|\sigma_r - \sigma_z| / \sqrt{3}}{(2\sigma'_r + \sigma'_z)/3} \quad \text{and} \quad \sin\phi = \frac{3f}{2\sqrt{3} + f} \tag{1}$$

where $\sigma'_z$ and $\sigma'_r$ are respectively the Terzaghi effective axial and radial stresses:

$$\sigma'_r = \sigma_r - P_p \ , \sigma'_z = \sigma_z - P_p \qquad (2)$$

The corresponding stress-strain curves are shown on Fig. 6. The mobilized friction coefficient and friction angle are plotted on Fig. 7a,b.

For the test at 8 MPa confining pressure, the loading phase was interrupted at an axial strain of 7.2 % corresponding to a deviatoric stress of 9.9 MPa. This deviatoric stress is actually less than the stress at which failure occurs. Thus the same sample could be used for another experiment if loaded at a different higher pressure. For the tests at 16 MPa and 18 MPa, the material was compacting and reached the critical state (i.e. zero dilatancy) for a maximum friction angle of $\phi$=28°.

The softening and dilatant behaviour which occurs at the end of the test at 18 MPa of confinement is attributed to the occurrence of shear-banding in the sample.

## *2.2 2D-constitutive equations*

In view of analysing shear band formation inside the fault zone, the 2D-constitutive equations of Mohr-Coulomb flow elastoplasticity theory are used here. Considering only the components of the stress tensor ($\sigma_{ij}$) in the plane of deformation, the components of the 2D stress tensor are decomposed into a spherical and a deviatoric part:

$$\sigma_{ij} = s_{ij} + \sigma\delta_{ij} / 2 \qquad (3)$$

where $\sigma = \sigma_{kk} / 2$ is the 2D mean stress and $\delta_{ij}$ is the Kronecker index. The shearing stress intensity $\tau$ is defined as follows (square root of the second deviatoric stress invariant $J_{2s}$)

$$J_{2s} = s_{ij} s_{ij} / 2 \quad ; \quad \tau = \sqrt{J_{2s}} \qquad (4)$$

We use the usual decomposition of the strain tensor **ε** and its deviator **e** in elastic and plastic part:

$$\varepsilon_{ij} = \varepsilon_{ij}^{e} + \varepsilon_{ij}^{p} \; ; \; e_{ij} = e_{ij}^{e} + e_{ij}^{p} \qquad (5)$$

The elastic strain rates $\dot{\varepsilon}_{ij}^{e}$ are related to the stress rates through the equation of linear isotropic elasticity

$$\dot{e}_{ij}^{e} = \dot{s}_{ij} / 2G \; ; \; \dot{\varepsilon}^{e} = \dot{\varepsilon}_{kk}^{e} = \dot{\sigma} / K \qquad (6)$$

where G and K are the 2D elastic shear and compression moduli respectively.

The yield function F and plastic potential Q are of the Coulomb type

$$F = \tau + \sigma f(\psi) \; ; \; Q = \tau + \sigma d(\psi) \qquad (7)$$

where $f(\psi)$ and $d(\psi)$ are the corresponding mobilized friction and dilatancy coefficients, which are both assumed to be functions of an appropriate hardening parameter $\psi$ (e.g. the accumulated plastic shear strain).

In the coordinate system of principal axes of initial stress, the incremental constitutive equations are (see Vardoulakis and Sulem, 1995 for details):

$$\begin{aligned} \Delta\sigma_{11} &= L_{11}\Delta\varepsilon_{11} + L_{12}\Delta\varepsilon_{22} \\ \Delta\sigma_{22} &= L_{21}\Delta\varepsilon_{11} + L_{22}\Delta\varepsilon_{22} \\ \Delta\sigma_{12} &= 2G\Delta\varepsilon_{12} \end{aligned} \qquad (8)$$

where for the considered 2D-constitutive model

$$
\begin{aligned}
L_{11} &= G\left(1+\kappa-\frac{1}{H}(1+\kappa f)(1+\kappa d)\right)\\
L_{12} &= G\left(-1+\kappa-\frac{1}{H}(1+\kappa f)(-1+\kappa d)\right)\\
L_{21} &= G\left(-1+\kappa-\frac{1}{H}(-1+\kappa f)(1+\kappa d)\right)\\
L_{22} &= G\left(1+\kappa-\frac{1}{H}(-1+\kappa f)(-1+\kappa d)\right)
\end{aligned}
\qquad (9)
$$

In equations (10), $\kappa = K/G = 1/(1-2\nu)$ ($\nu$ is the Poisson's ratio), $H = 1 + h$ is the plastic modulus ($h = df/d\psi$ is the hardening modulus).

## 3 STRAIN LOCALISATION ANALYSIS

In the following, the mechanism of shear band formation inside the clayey core is analyzed. As mentioned in the introduction, a clear shear plane is visible (Fig. 5). The striation indicates a slip at the centimetric scale (Fig. 5b,c). This plane is oriented with an inclination of 22° with respect to the horizontal axis. The angle between this slip plane and the fault plane is thus 38°. Extensive presentation of shear band analysis in geomaterials can be found in Vardoulakis and Sulem (1995). The mathematical details of the analysis are presented in Appendix.

The strain localisation analysis consists in searching the incipient of a shear band in a solid as a mathematical bifurcation condition for the deformation field. Considering an infinitesimal neighbourhood of a point in an elastic-plastic solid which is homogeneous as for the constitutive law and stress state, the strain localisation phenomenon is understood as the appearance of a discontinuity in strain rates which marks the onset of non-uniform response. Such a bifurcation of the velocity gradient along a loading path can be caused by material destabilising effects such as softening and lack of plastic normality in the constitutive law, as well as geometrical destabilising effects such as large deformation affecting equilibrium equations. This bifurcation condition is obtained from (a) the constitutive relationships of the material, (b) the conditions of mechanical equilibrium across the shear band boundary and (c) the kinematical compatibility conditions which expresses that the velocity field is to be continuous (no material discontinuity). The latter condition implies that only the normal component of the velocity gradient across the shear band is discontinuous whereas the tangential one is continuous (weak discontinuity). The above conditions describe the so-called 'continuous' bifurcation modes. It has been shown that the critical state for continuous bifurcation precedes the one for 'discontinuous' bifurcation where a discontinuity of the velocity field itself (and not only its gradient) is considered (Simo et al. 1993). Non-trivial solution for the condition of continuous bifurcation is a necessary condition for the shear band existence and provides both the shear band orientation and the deformation jump across the shear band. Rudnicki and Rice (1975) and Rice (1976) have obtained solutions for realistic elasto-plastic constitutive relationships for geomaterials.

For the 2D elasto-plastic constitutive equations (10), the localisation condition can be expressed explicitly in terms of the critical hardening modulus $h_B$ at shear banding and the orientation $\theta_B$ of the shear band (the details of the derivations are given in the Appendix)

$$h_B = \frac{(f-d)^2}{8(1-\nu)} \tag{10}$$

$$\theta_B = \pm\arctan\left(\frac{2(1-\nu)(2+f+d)^2}{(1-2\nu)(f^2+d^2)+2(3-2\nu)df+8(1-\nu)(1-2\nu)}\right)^{1/4} \tag{11}$$

As already pointed out by Rudnicki and Rice (1975) and Vardoulakis and Sulem (1995), equation (10) suggests that shear banding in plane strain deformation always takes place in the hardening regime ($h_B > 0$).

In triaxial tests, it is obtained that shear banding is occurring at the limit state of maximum friction coefficient (i.e. $f = \sin\phi$, $\phi = 28^\circ$, h=0) and at critical state (i.e. d=0). It is expected that in 2D conditions the same conditions are met. On Figure 8, the critical hardening modulus $h_B$ is plotted versus the dilatancy coefficient d for f=0 and for different values of the Poisson's ratio $\nu$ (0.2, 0.25, 0.3).

The shear band orientation is plotted on Figure 9a versus the dilatancy coefficient d for various values of the Poisson's ratio and on Figure 9b versus the Poisson's ratio for various values of the dilatancy coefficient.

## 4 DISCUSSION

The above analysis show that depending on the Poisson's ratio and on the dilatancy angle of the gouge the orientation angle of the shear band is between 50° and 55° with respect to the principal stress axes. It is reasonable to assume that shear banding occurs in 2D at critical state (d=0) as observed in triaxial test. For a value of 0.2 for the Poisson's ratio, the orientation of the shear band is 52°. This result compares very well with the so-called Arthur solution $\theta_B = 45° + \phi/4 + \psi/4$ where $\psi$ is the dilatancy angle (equal to zero at shear banding here) (Arthur et al. 1977, Vardoulakis 1980, Vardoulakis and Sulem, 1995). Note that the classical Coulomb solution for the orientation of the failure surface is $45°+\phi/2= 59°$. The Coulomb solution states that failure occurs at the state of maximum mobilized friction and that the orientation of the failure plane coincides with the planes across which the ratio of shear to normal stress is maximum. The bifurcation solution derived from the above shear band analysis accounts for the volumetric behaviour of the material. In case of associative plasticity, the Coulomb and the Arthur solutions coincide.

From this result and from the observation that the slip surface is oriented at an angle of about 22° with respect to the horizontal axis, we deduce that the principal stress axes are rotated with between an angle comprised between 28° and 33° with respect to the horizontal and vertical axes. They correspond approximately to the directions normal and parallel to the major fault (oriented at 60°) (Fig. 10).

This result should be corroborated with in situ direct stress measurements which are not available for the moment. It is compatible with the general observation that in weak fault zones the fault actually support very little shear stress like for example the San Andreas Fault (California) (Lachenbruch and Sass 1992). It confirms the general observation that slip lines might be influenced by very local conditions (here the clayey core of the fault) and the local movements can geometrically differ from what is predicted from large scale observations. Significant stress rotation within fault zones has been observed because of fault slip (Fitzens and Miller, 2004). It is also commonly inferred in weak faults (Faulkner et al 2006). In Aigion fault, the presence of a very narrow zone (one meter thick) of soft clayey material with a very low permeability ($10^{-19}m^2$, after Sulem et al. 2004) and a lower friction angle (28°) than the surrounding limestone (45°), can explain the stress rotation. Due to the very low permeability of the fault, this zone acts as an impervious barrier to fluid flow and allows fluid pressurisation during a seismic event (Rice, 2006, Sulem et al 2006). Stress rotation can be seen as a consequence of high pore fluid pressure trapped inside the fault core and effective mean stress decrease. The stress rotation is also triggered by the reduction of the friction angle inside the fault core and Coulomb plasticity effect (Lockner and Byerlee, 1993).

The understanding of the stress field surrounding active fault systems is important for seismic risk assessment. In particular, a key question is how long a seismic slip on a fault perturbs the surrounding stress field. High pore pressure propagation inside surrounding damage zones can make the link between earthquakes and after shocks.

## Acknowledgements

The authors wish to acknowledge the EU projects *DG-Lab Corinth* (EVR1-CT-2000-40005) and *Fault, Fractures and Fluids: Golf of Corinth*, in the framework of program Energy (ENK6-2000-0056) and the 'Groupement de Recherche Corinthe' (GDR 2343), INSU-CNRS for supporting this research.

**APPENDIX**

**Shear band analysis for plane strain deformation**

As the reference coordinates system we consider in the following the principal axes of the stress tensor ($X_1$, $X_2$). Strain localization in the form of a shear band corresponds to weak discontinuities for the incremental displacement and occurs first when the acoustic tensor $\mathbf{\Gamma}$ becomes singular (Hill, 1962):

$$\det(\Gamma_{ij}) = 0, \text{ with } \Gamma_{ij} = L_{ikjl} n_k n_l \tag{A1}$$

where $\mathbf{L}$ is the incremental constitutive tensor of the material and $n_i$ are the cosines of the normal to the shear band.

Using the 2D constitutive equations equations (8) the acoustic tensor $\mathbf{\Gamma}$ is expressed as

$$\mathbf{\Gamma} = \begin{pmatrix} L_{11} n_1^2 + G n_2^2 & (L_{12} + G) n_1 n_2 \\ (L_{21} + G) n_1 n_2 & L_{22} n_2^2 + G n_1^2 \end{pmatrix} \tag{A2}$$

and its determinant as a fourth order polynomial of the cosines of the normal to the shear band $n_i$

$$\det(\mathbf{\Gamma}) = G L_{11} n_1^4 + (L_{11} L_{22} - L_{12} L_{21} - G L_{12} - G L_{21}) n_1^2 n_2^2 + G L_{22} n_2^4 \tag{A3}$$

This leads to the characteristic equation for the shear band inclination angle $\theta$

$$a \tan^4 \theta + b \tan^2 \theta + c = 0 \tag{A4}$$

where θ is measured with respect to minor (in absolute value) principal stress direction

$$\tan\theta = -(n_1 / n_2) \tag{A5}$$

and

$$\begin{aligned} a &= L^*_{11} \\ b &= L^*_{11}L^*_{22} - L^*_{12}L^*_{21} - L^*_{12} - L^*_{21} \\ c &= L^*_{22} \\ \text{for } i &= 1,2 \;\; L^*_{ij} = L_{ij} / G \end{aligned} \tag{A6}$$

The condition for shear band formation is derived from the requirement that the characteristic equation (A4) has real solutions. This condition is firstly met at a state $C_B$ (B for bifurcation) for which

$$b / a < 0 \quad \text{and} \quad D = b^2 - 4ac = 0 \tag{A7}$$

For any state beyond $C_B$, there are four solutions for the shear-band orientation. According to the observations, the shear bands usually belong to a single family of symmetric solutions. This statement justifies the selection of (A7) as the shear-band bifurcation condition. At $C_B$, only two symmetric shear-band directions exist, given by

$$\theta_B = \pm\arctan\left(c^{1/4}\right) \tag{A8}$$

Using the constitutive equations (9), equation (A7) can be solved explicitly in terms of the critical hardening modulus $h_B$ at shear banding

$$h_B = \frac{(f-d)^2}{8(1-\nu)} \tag{A9}$$

As already pointed out by Rudnicki and Rice (1975) and Vardoulakis and Sulem (1995), equation (A9) suggests that shear banding in plane strain deformation always takes place in the hardening regime ($h_B > 0$).

From equations (A8) and (A9) we obtain:

$$\theta_B = \pm \arctan\left( \frac{2(1-\nu)(2+f+d)^2}{(1-2\nu)(f^2+d^2)+2(3-2\nu)df+8(1-\nu)(1-2\nu)} \right)^{1/4} \tag{12}$$

The shear band orientation is plotted on Figure 9a versus the dilatancy coefficient d for various values of the Poisson's ratio and on Figure 9b versus the Poisson's ratio for various values of the dilatancy coefficient.

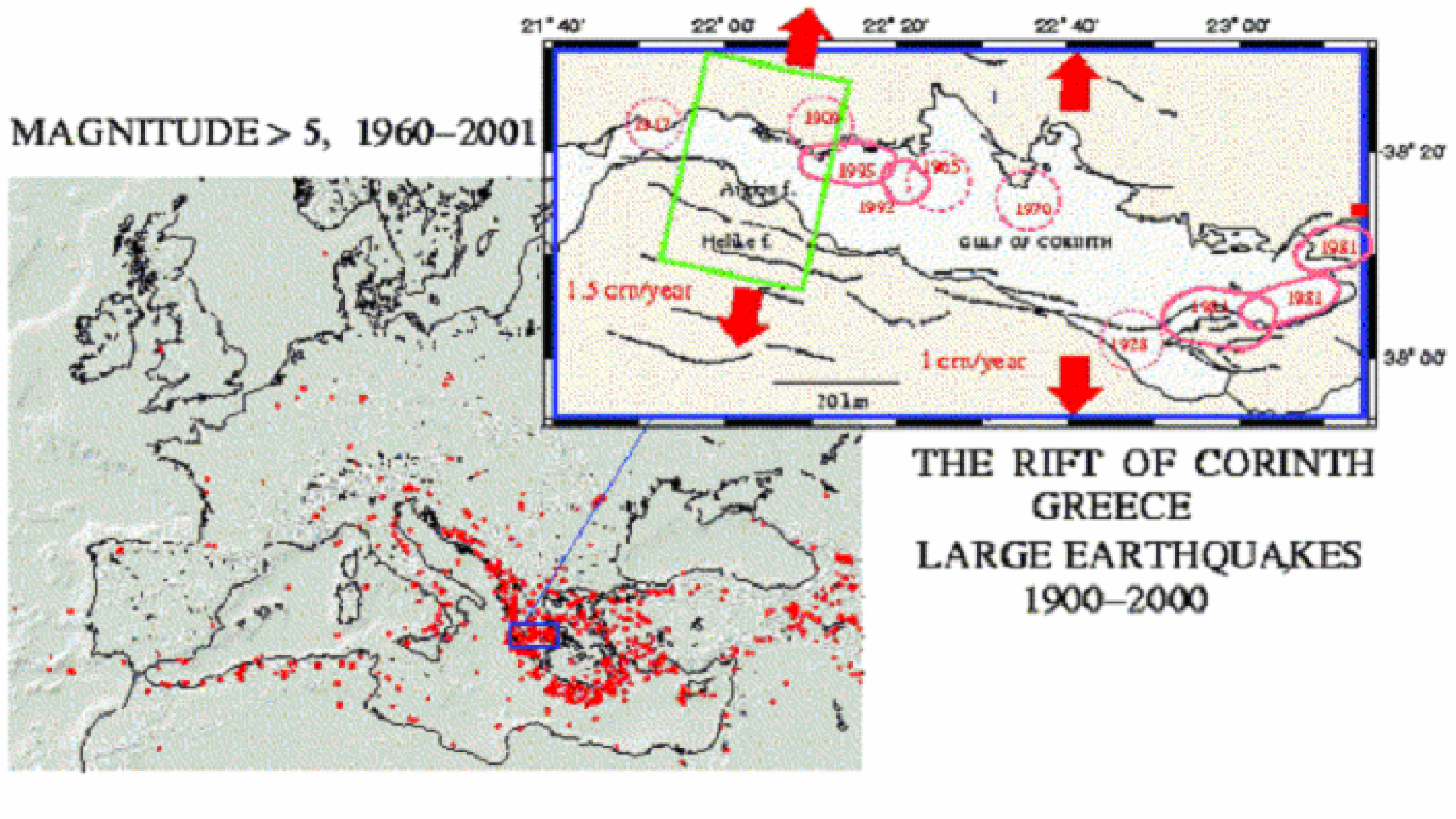

***Fig.1. General sismo-tectonic context of the Corinth Rift. The CRL location is the green rectangle centred on Aigion. Recent earthquakes are indicated by their date and the approximate size of their seismic rupture. The European map shows all seismic events with magnitude larger than 5, as observed from 1960 to 2001. (from Cornet et al, 2004a)***

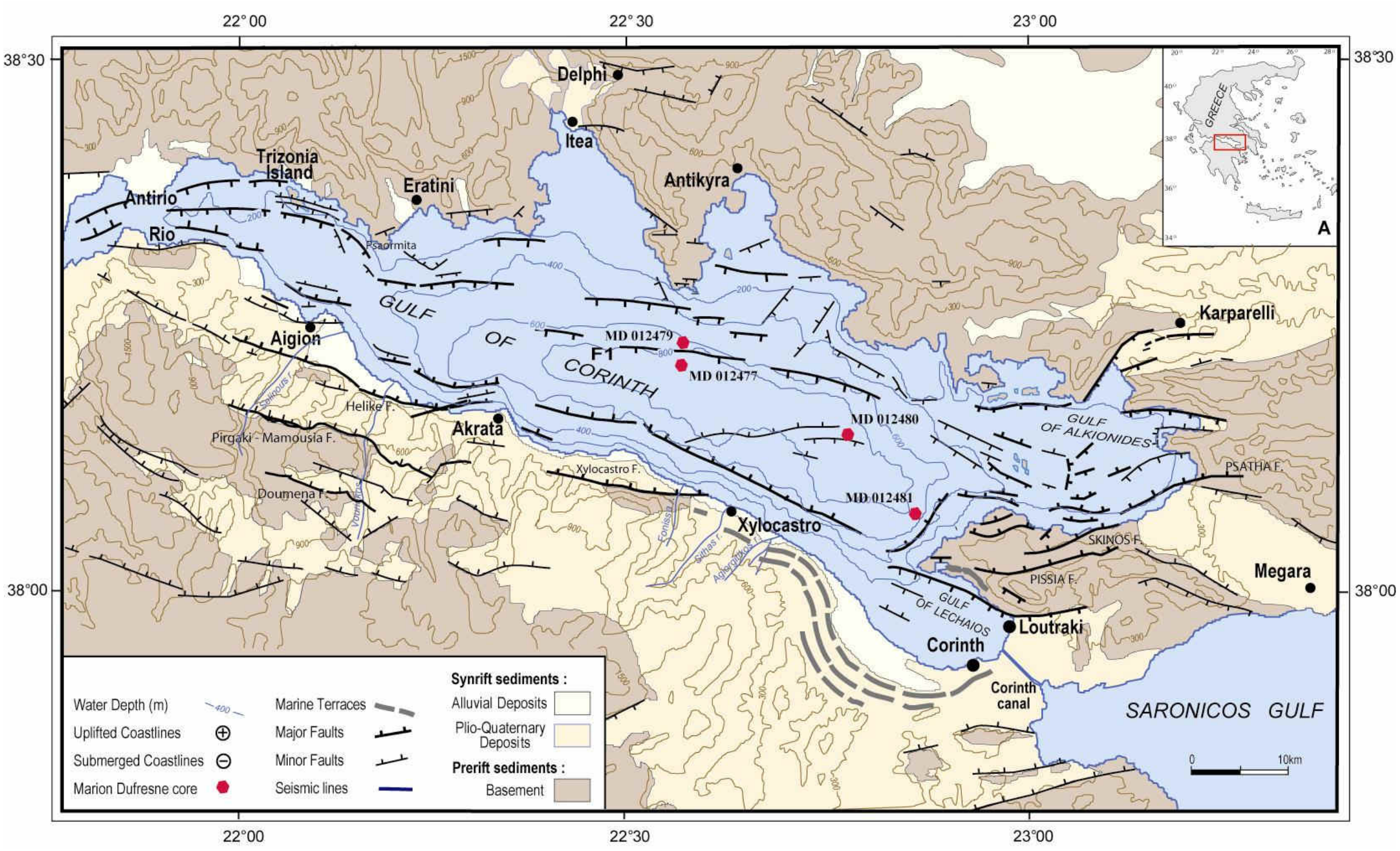

***Fig. 2 Geological map of the Gulf of Corinth (from Moretti et al, 2003)***

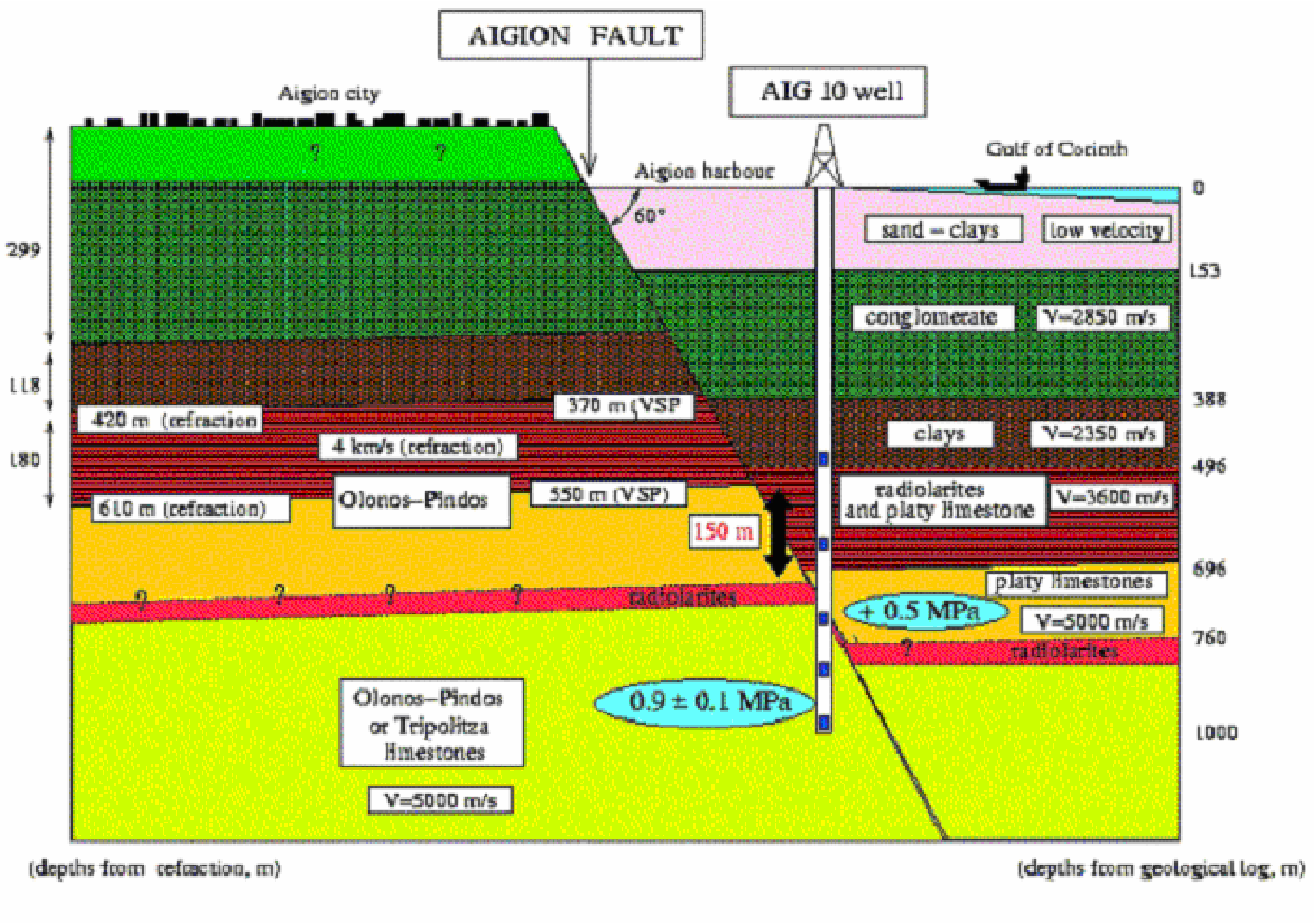

**Fig. 3 : Schematic structural cross section through Aigion Fault, in Aigion harbour (from Cornet et al, 2004b).**

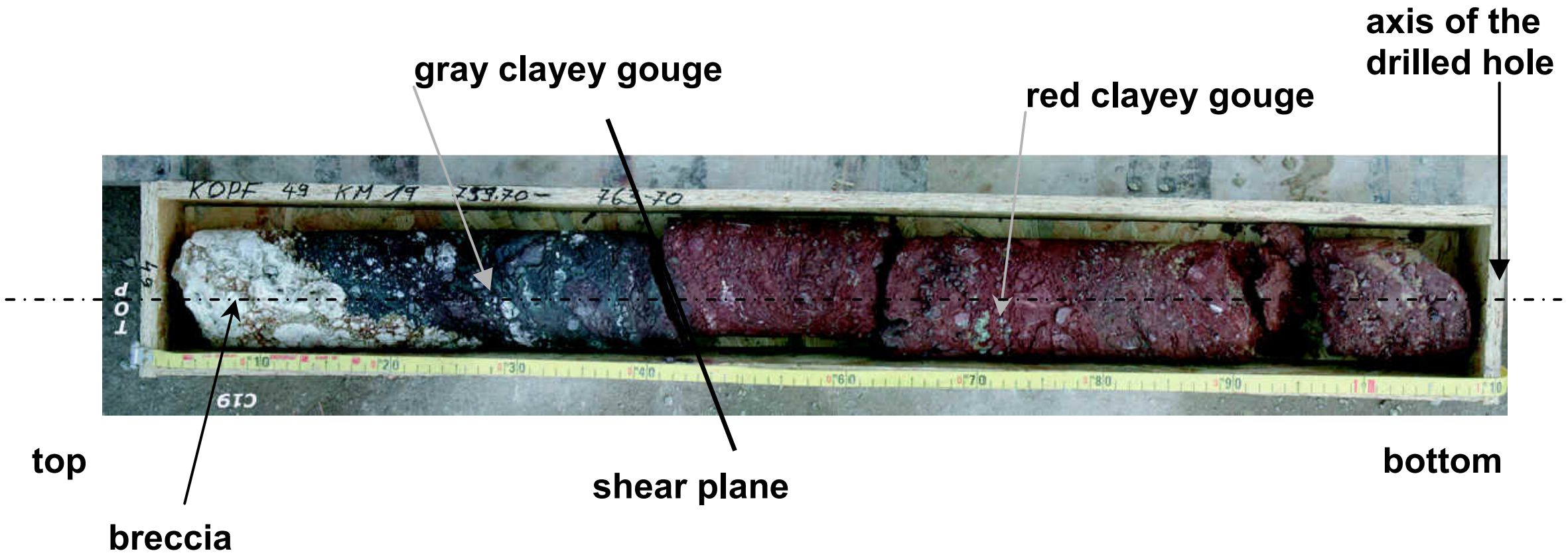

***Fig. 4: Box 49 containing the core at depth 759.70m, characterized as "Aigion fault" core***

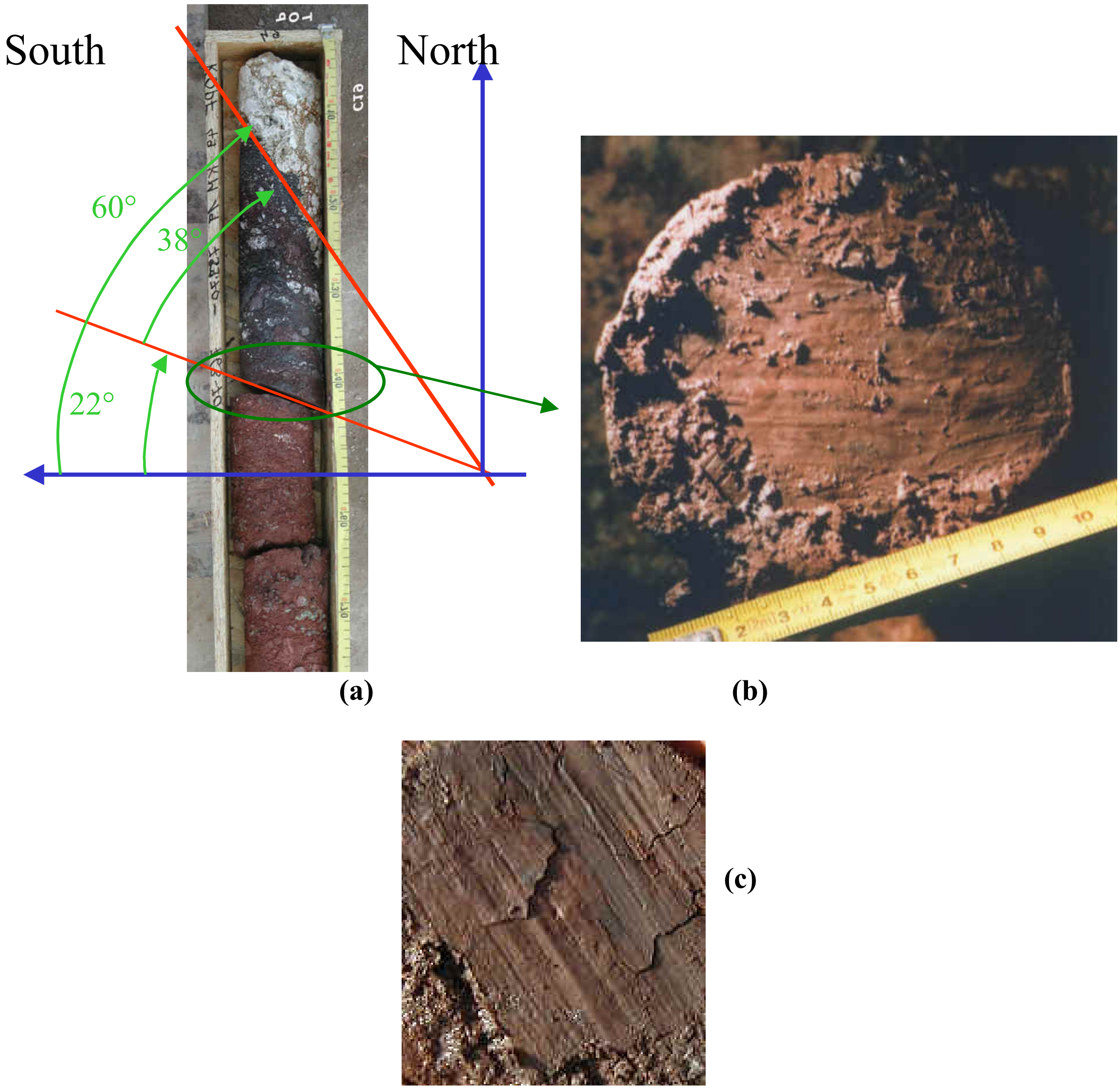

**(a)** **(b)**

**(c)**

***Fig. 5 : (a) The clayey core retrieved from Aigion fault ; (b and c) Striation of the slip plane***

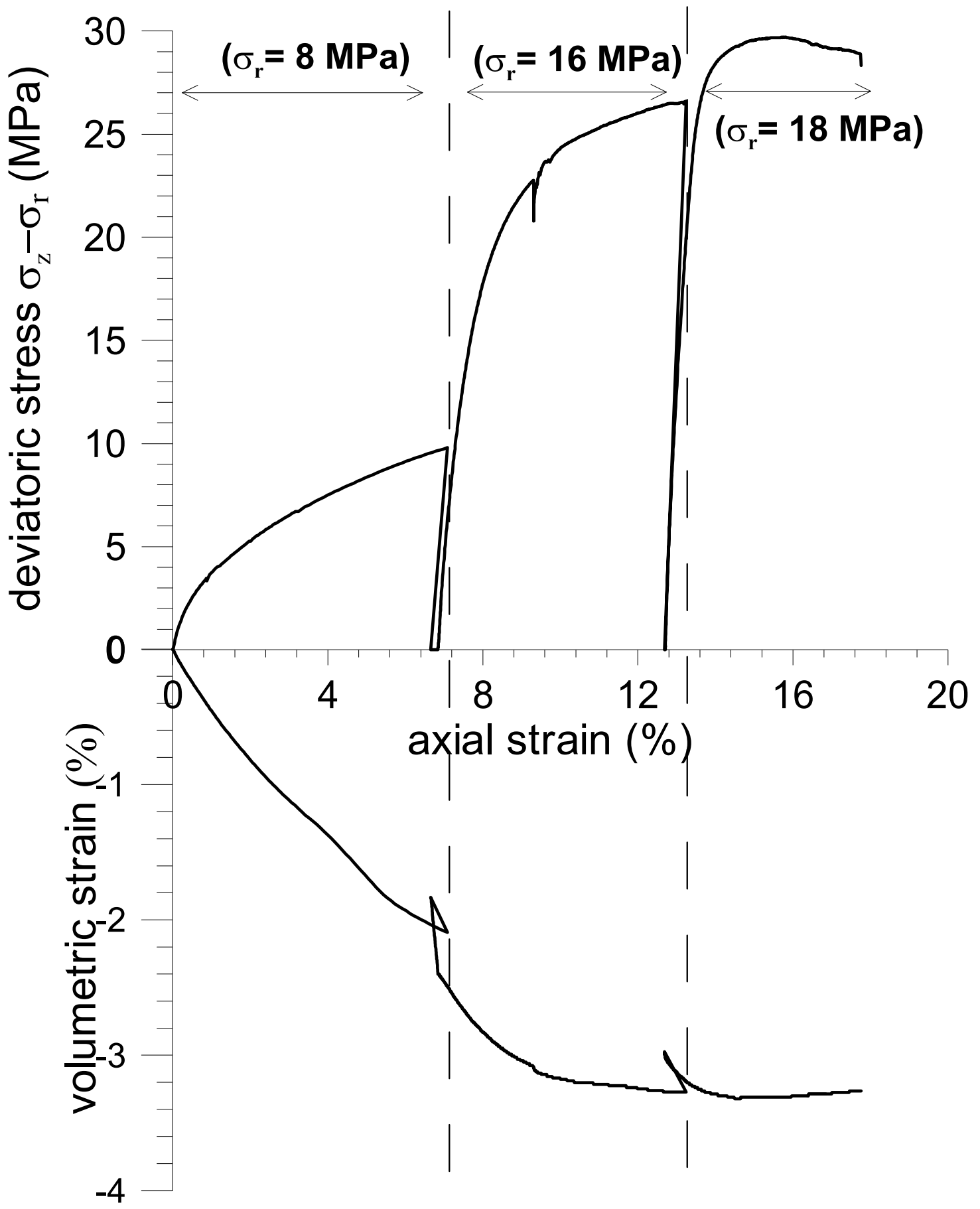

***Fig.6 : Drained triaxial tests at room temperature (22°C)***

.

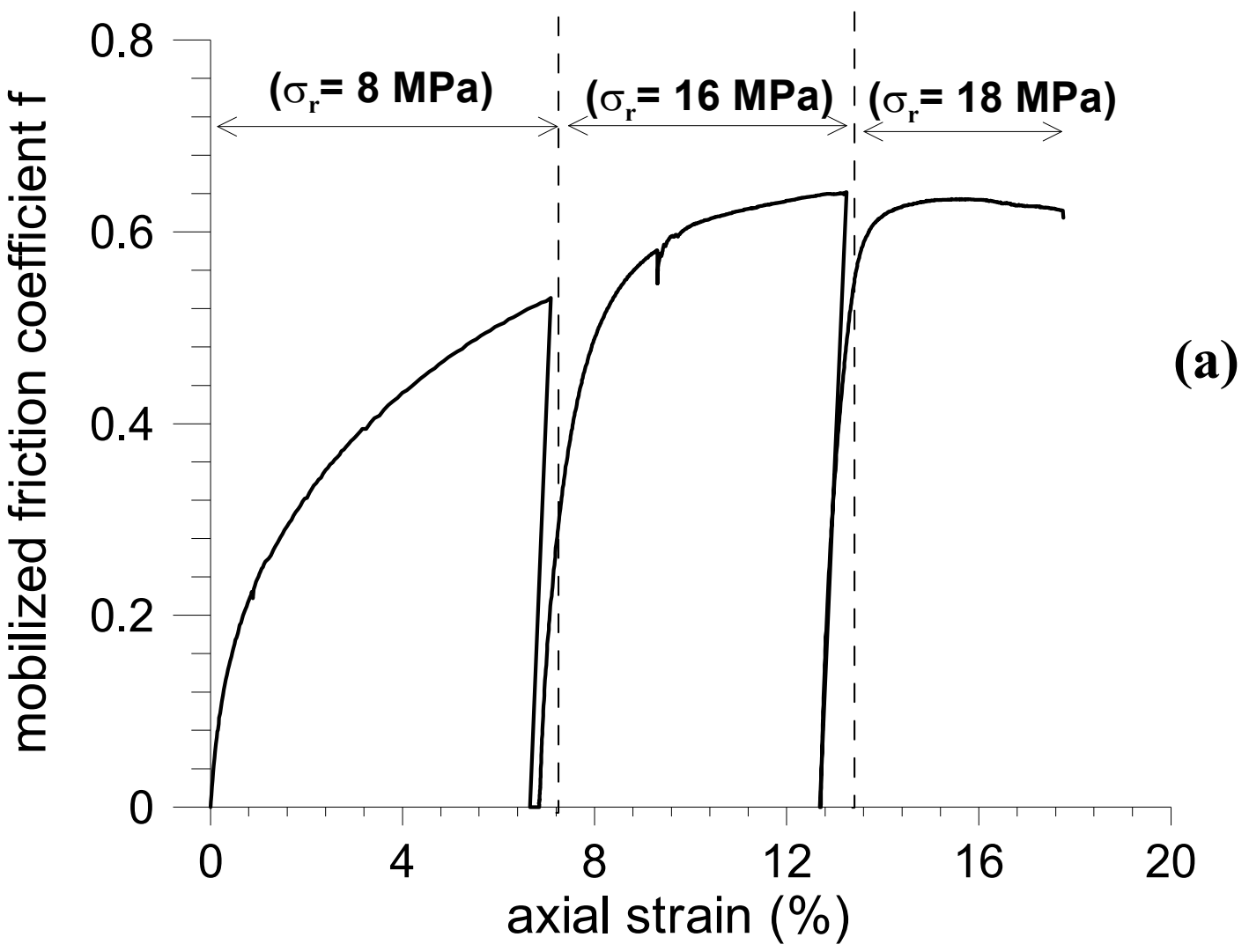

***Fig.7 : Mobilized friction on drained triaxial tests at room temperature (22°C)***

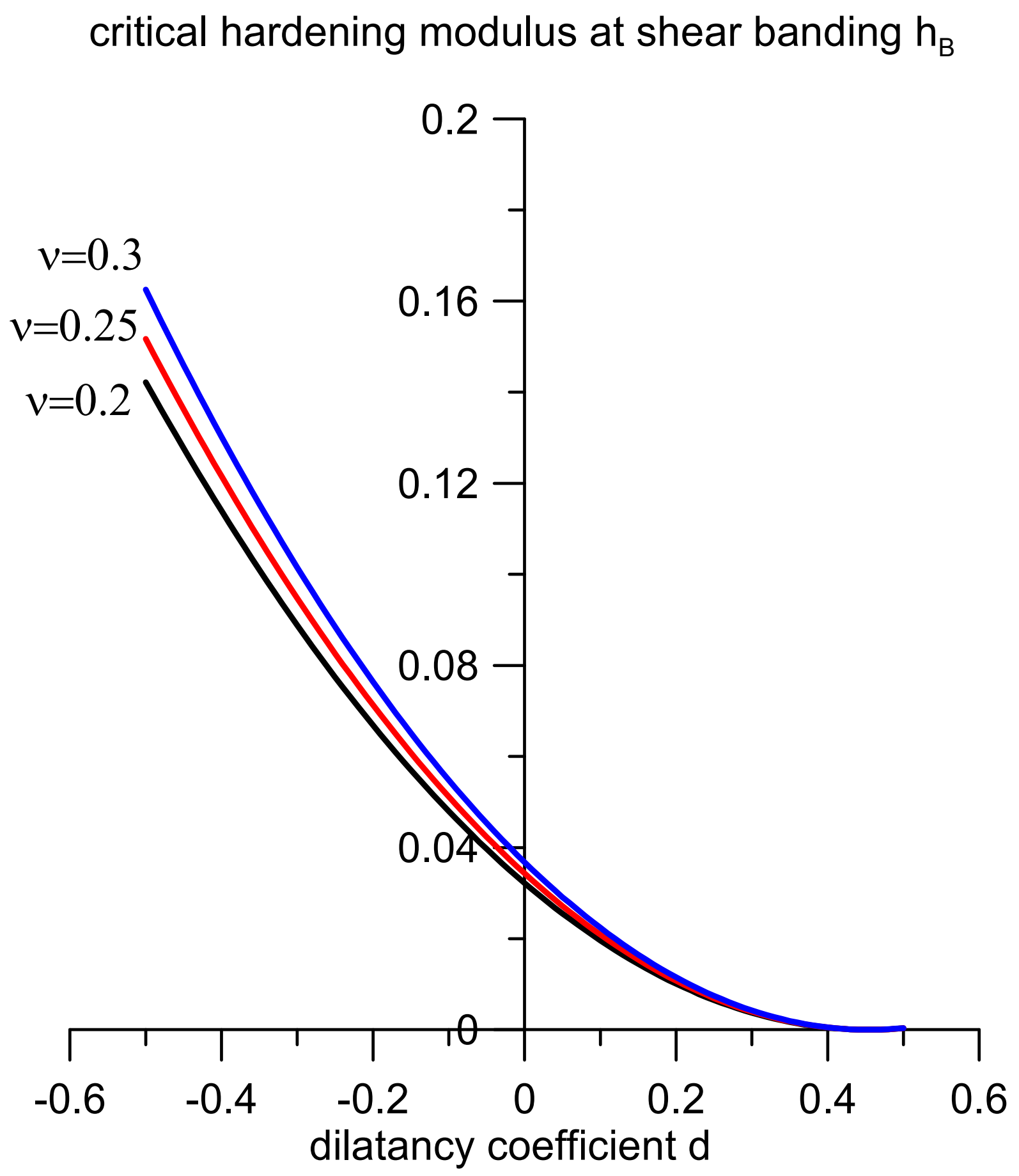

***Fig. 8: Critical hardening modulus at shear banding***

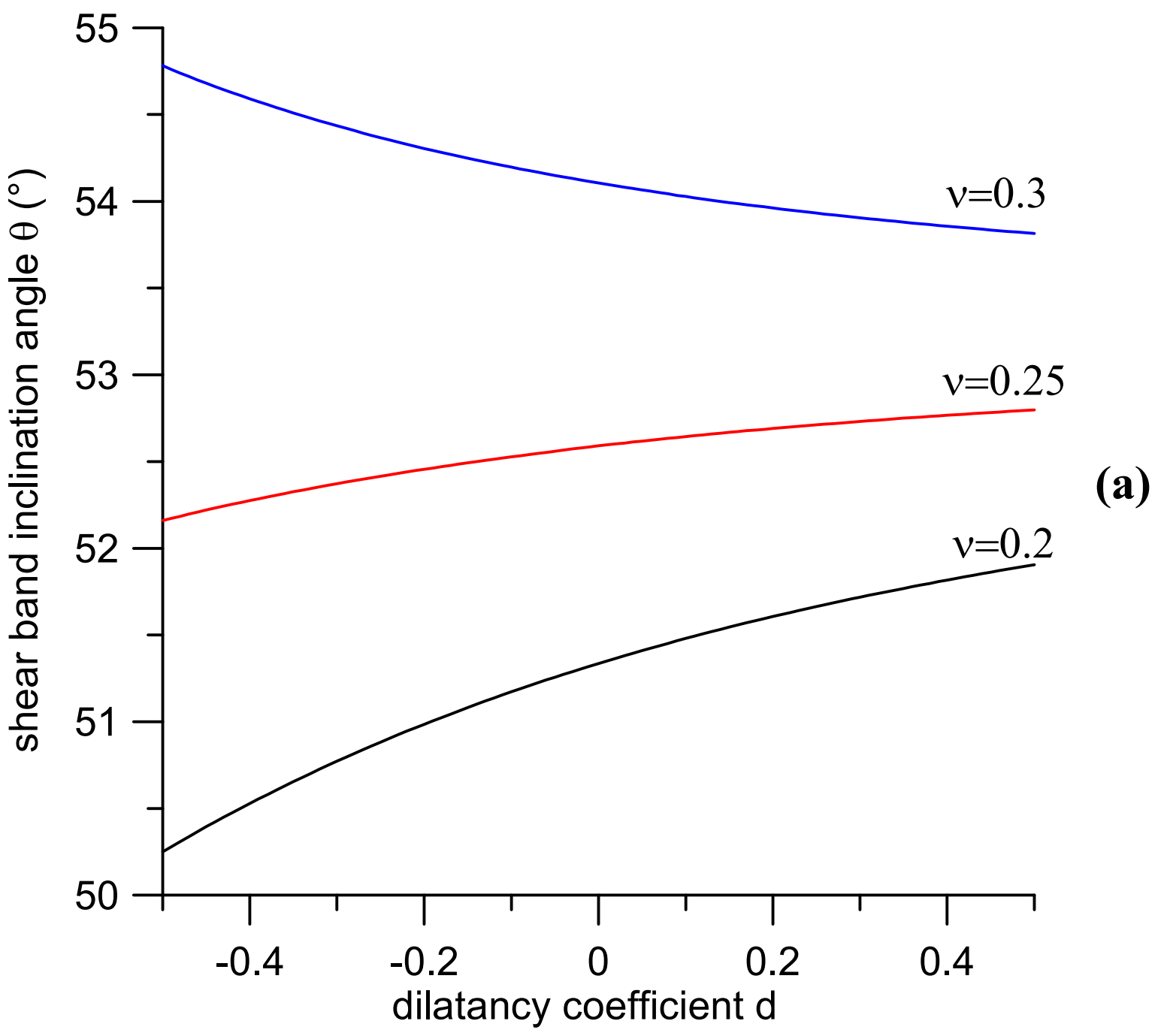

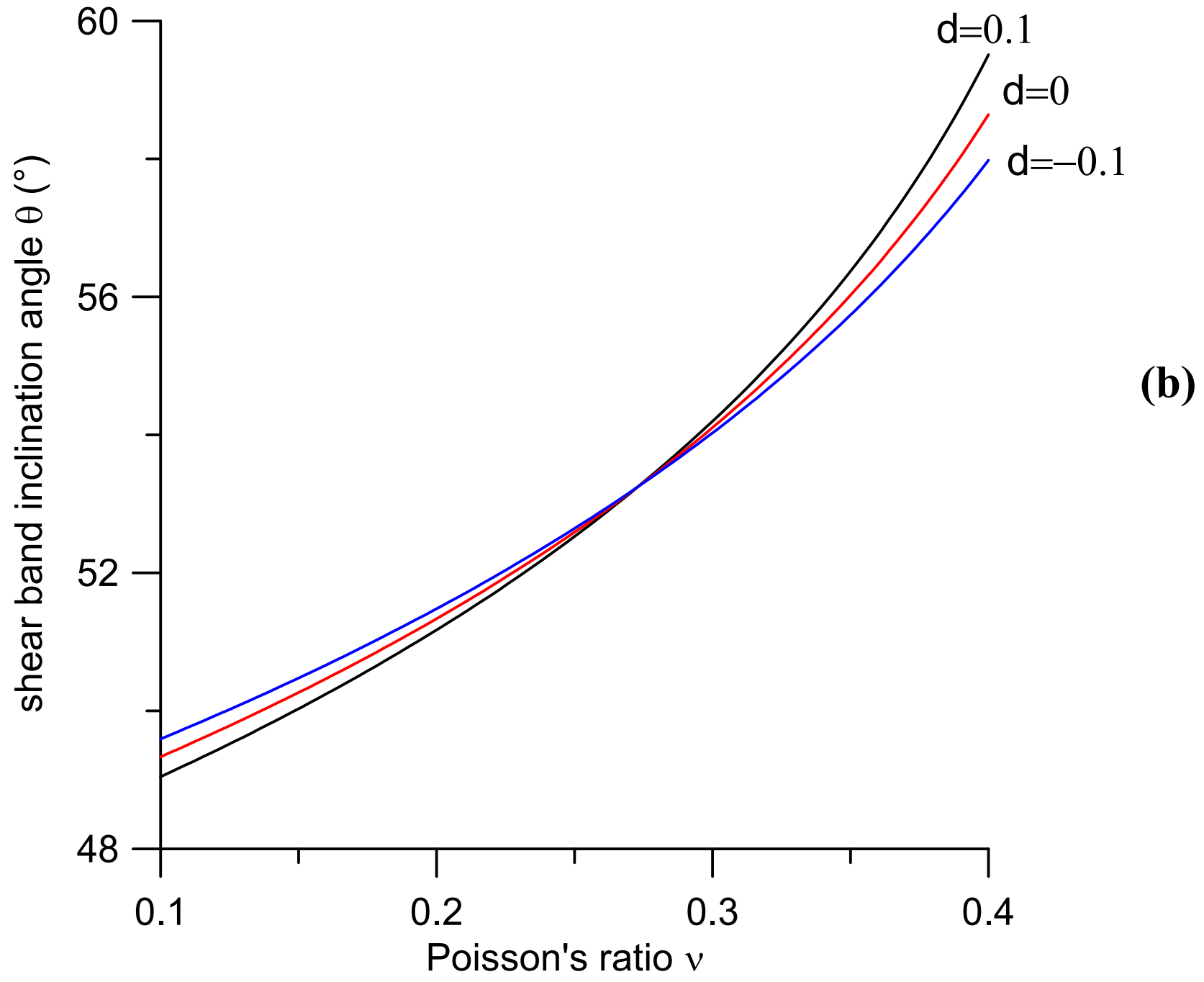

***Fig. 9: Shear band inclination angle in plane strain: (a) effect of dilatancy coefficient, (b) effect of Poisson's ratio***

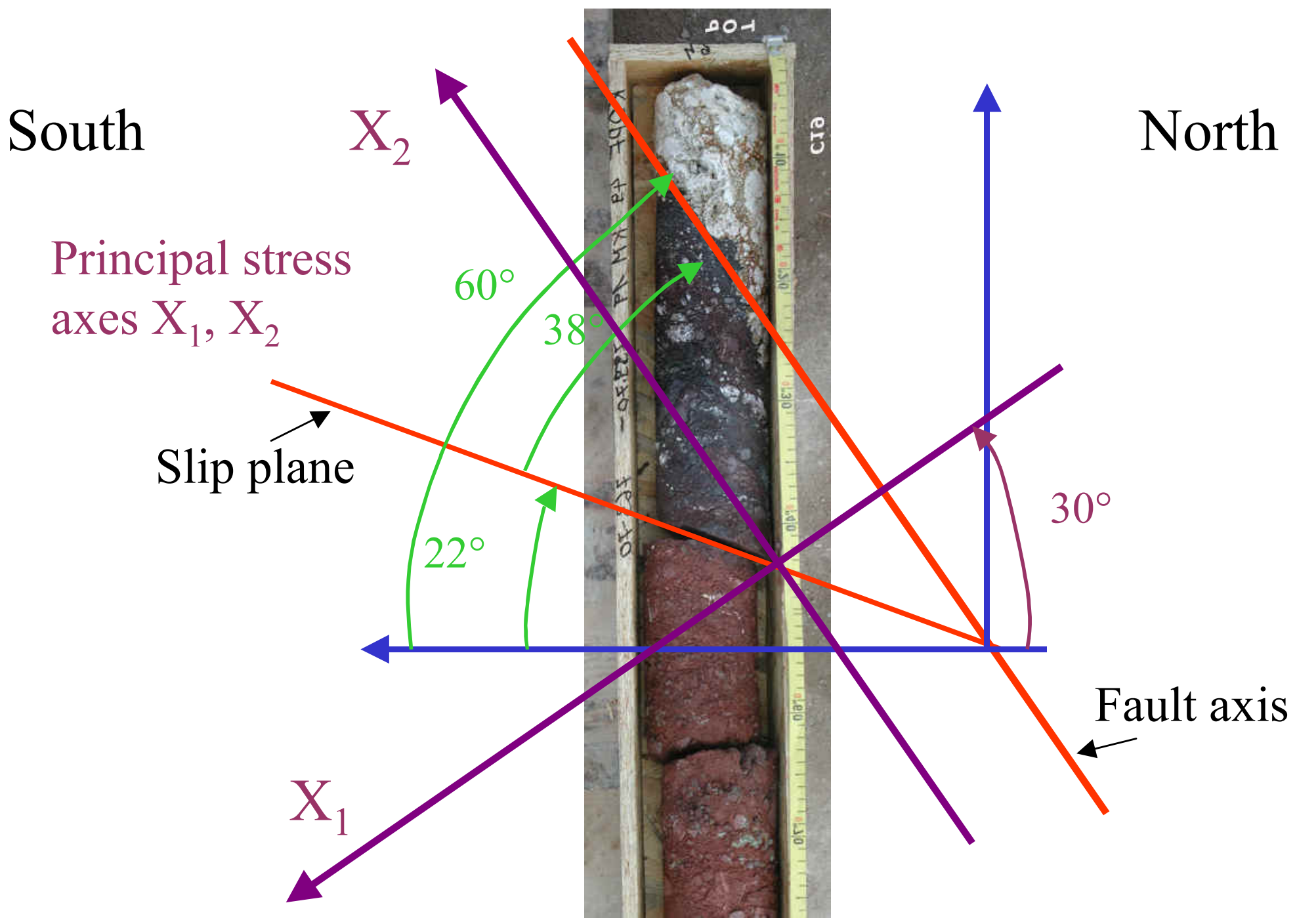

***Fig. 10: Principal stress axes as deduced from shear band analysis: $X_1$ is the minor (in absolute value) principal axis***